\documentclass[twocolumn,amsmath,amssymb,prb,longbibliography]{revtex4-1}
\usepackage{multirow}
\usepackage{array}
\usepackage{amsmath}
\usepackage{graphicx}
\usepackage{dcolumn}
\usepackage{bm}
\usepackage{verbatim}
\DeclareMathAlphabet \mathbfcal{OMS}{cmsy}{b}{n}
\begin{document}




\title{Topological resonance and single-optical-cycle valley polarization in gapped graphene}

\author{S. Azar Oliaei Motlagh}
\author{Fatemeh Nematollahi}
\author{Vadym Apalkov}
\author{Mark I. Stockman}
\affiliation{Center for Nano-Optics (CeNO) and
Department of Physics and Astronomy, Georgia State
University, Atlanta, Georgia 30303, USA
}

\date{\today}
\begin{abstract}

For gapped graphene, we predict that an intense ultrashort (single-oscillation) circularly-polarized optical pulse can induce a large population of the conduction band and a large valley polarization. With an increase in the bandgap, the magnitude of the valley polarization gradually increases from zero (for the native gapless graphene) to a value on the order of unity. The energy bandwidth of the electrons excited into the conduction band can be very large ($\gtrsim 10$ eV for a reasonable pulse amplitude of $\sim 0.5$ $\mathrm{V/\AA}$). These phenomena are due to the effect of topological resonance:  the matching of the topological (geometric) phase and the dynamic phase. Gapped graphene with tunable bandgap can be used as a convenient generic model of two-dimensional semiconductors with honeycomb generic lattice structures and broken inversion symmetry, such as transition metal dichalcogenides.

\end{abstract}
\maketitle
\section{Introduction}


Interactions of ultrafast intense optical pulses with solids create a unique platform to study highly nonlinear phenomena such as ultrafast field driven currents, high harmonic generation, and ultrafast ionization \cite{Schiffrin_at_al_Nature_2012_Current_in_Dielectric, Apalkov_Stockman_PRB_2012_Strong_Field_Reflection, Higuchi_Hommelhoff_et_al_Nature_2017_Currents_in_Graphene, Gruber_et_al_ncomms13948_2016_Ultrafast_pulses_graphene, Stockman_et_al_PhysRevB.95_2017_Crystalline_TI,Stockman_et_al_PhysRevB.98_2018_3D_TI,Stockman_et_al_PhysRevB.99_2019_Weyl, Hommelhoff_et_al_PhysRevLett.121_2018_Coherent, Ghimire_et_al_Nature_Communications_2017_HHG, Reis_et_al_Nat_Phys_2017_HHG_from_2D_Crystals, Simon_et_al_PRB_2000_Strong_Field_Fs_Ionization_of_Dielectrics, Hommelhoff_et_al_1903.07558_2019_laser_pulses_graphene, sun_et_al_nnano.2011.243_2012_Ultrafast_pulses_graphene, Mashiko_et_al_Nature_Communications_2018_ultrafast_pulse_solid, Shin_et_al_IOP_Publishing_2018_ultrafast_pulse_solid}. 
Previously several theoretical and experimental works have addressed the problems of ultrafast nonlinear electron dynamics in graphene and graphene-like materials, including  monolayers of transition metal dichalcogenides (TMDC's) \cite{Hommelhoff_et_al_PhysRevLett.121_2018_Coherent, 
Gruber_et_al_ncomms13948_2016_Ultrafast_pulses_graphene, Higuchi_et_al_Nature_2017, Leitenstorfer_et_al_PhysRevB.92_2015_Ultrafast_Pseudospin_Dynamics_in_Graphene, Stockman_et_al_PhysRevB.96_2017_Berry_Phase, Stockman_et_al_PhysRevB.98_2018_Rapid_Communication_Topological_Resonances, Sun_et_al_Chinese_Physics_B_2017_Ultrafast_pulses_TMDC, Zhang_et_al_OSA_2018_ultrafast_pulse_TMDC}. 

Graphene is a two-dimensional (2d) crystal with unique physical properties \cite{Novoselov_Geim_et_al_nature04233_2D_Electrons_in_Graphene, Novoselov_et_al_Nature_Materials_2007_Rise_of_Graphene, Electronic_properties_graphene_RMP_2009, Kane_Mele_PhysRevLett.95_Spin_Hall_Effect_in_Graphene}. 
Pristine graphene has honeycomb crystal structure with two sublattices, $A$ and $B$ -- see Fig.\ \ref{fig:Energy} below in Sec.\ \ref{Model_and_Equations}. It possesses both inversion ($\mathcal P)$ and time-reversal ($\mathcal T$) symmetries. Accordingly, in the reciprocal space, there are two inequivalent points and the corresponding valleys, $K$ and $K^\prime$.
The energy dispersion near these $K$- or $K^\prime$-points is described by a massless Dirac Hamiltonian. 

Due to the inversion ($\mathcal P$) symmetry, the electron dispersion in graphene is gapless. The valley polarization induced by ultrashort chiral (i.e., with fields rotating) laser pulses is rather small \cite{Stockman_et_al_PhysRevB.93.155434_Graphene_Circular_Interferometry}. In contrast, if a honeycomb lattice is not center-symmetric, then the electron spectrum has an energy gap and the valley polarization induced by chiral single-oscillation pulses may be rather large, as we have predicted for TMDC's \cite{Stockman_et_al_PhysRevB.98_2018_Rapid_Communication_Topological_Resonances}. A natural question arises: does the single-cycle valley polarization stems just from a chemical difference between graphene and TMDC's (in particular, the presence of a transition metal), an appreciable spin-orbit interaction inherent in TMDS's but not in graphene, or simply due to a presence of the bandgap? 

In this Article we show that in the gapped graphene, which is chemically identical to the pristine graphene and whose inversion symmetry is broken by the external environment, the chiral optical single-oscillation optical pulses can induce extremely high valley polarization. We attribute such a high magnitude of the valley polarization to the effect of topological resonance   \cite{Stockman_et_al_PhysRevB.98_2018_Rapid_Communication_Topological_Resonances}, which appears in TMDC's and is due to the mutual compensation of the dynamic phase, $\Delta_g t$, where $\Delta_g$ is the bandgap, and $t$ is time, and the topological (Berry) phase. Such a compensation can take place for a TMDC's and other 2d honeycomb crystals with a broken $\mathcal P$ symmetry but not in graphene, which is $\mathcal P$-symmetric and where  $\Delta_g=0$, which explains this difference between graphene and TMDC's.

The symmetry between the $K$- and $K^\prime$ valley in  the 2d honecomb crystals is strictly protected by the time-reversal ($\mathcal T$) symmetry irrespective of the $\mathcal P$-symmetry. A linearly-polarized light causes valley currents, which causes valley separation and breaks the valley symmetry in real space. \cite{Golub_et_al_PRB_2011_Valley_Separation_by_Polarized_Light} In contrast, to generate the valley polarization (i.e., asymmetry in the valley populations, which is uniform in the real space), it is necessary to break the $\mathcal T$-symmetry, which can be done with circularly polarized radiation, magnetic fields, or spin-polarized carriers. The optical generation of the valley polarization in gapped graphene in a regime of relatively weak fields and a continuous-wave (CW) excitation with circularly polarized light has been discussed in Refs. \onlinecite{Kibis_et_al_PhysRevB.95.125401_2017_gapped_Dirac, Rycerz_et_al_Nature_2007_gapped_graphene, Xiao_valley_Hall_PRL_2007, Niu_et_al_PhysRevB.77_2008_Valley_Selection_in_Graphene}. In the presence of a strong magnetic field or by injection of spin-polarized electrons, a high valley polarization can be achieved. \cite{Ye_et_al_Nature_Nanotechnology_2016_Electrical_generation_and_control,  Manchon_et_al_PRB_2016_Valley-Dependent_Spin-Orbit_Torques}
In the presence of the valley polarization, spatio-temporal symmetry of graphene is reduced allowing, in particular, for second harmonic generation. \cite{Golub_Tarasenko_PRB_2014_Valley_Polarization_SHG_in_Graphene} 

Previously, we predicted  \cite{Stockman_et_al_PhysRevB.98_2018_Rapid_Communication_Topological_Resonances} a strong valley polarization in the monolayer TMDC's, $\mathrm{MoS_2}$ and $\mathrm{ W S_2}$ which was induced by a single cycle of a chiral (``circularly polarized'') pulse \footnote{Following a tradition, we will be calling such pulses circularly polarized though for a single-oscillation chiral pulse the electric field cannot change simply rotate -- it has to change in magnitude from zero to its maximum amplitude and then back to zero during the pulse}. Here we aim to show that  the fundamentally fastest induction of the valley population in the 2d honeycomb  crystals does not fundamentally depend on either the spin-orbit coupling or chemical composition of the monolayers. Namely, we predict that it is possible to generate a high valley polarization in the gapped graphene with a chiral, ultimately-short  (single-oscillation) optical pulse whose duration is just a few femtoseconds with a realistic amplitude of $\sim 0.5 \mathrm{V/\AA}$. 

This is a non-resonant excitation, because the spectral width of such a pulse is on the order of its frequency. We will show below in this Article that the ultrafast induction of the high valley polarization is a fundamentally non-perturbative effect that is due to electron Bloch motion in the reciprocal space.  Due to a large amplitude of the optical pulse, the electron Bloch trajectory during the pulse extends over a significant part of the Brillioun zone. This causes electrons excited from the valence band (VB) into the conduction band (CB) to occupy  Bloch states in an extremely wide energy range, $\gtrsim 12$ eV. This enormous bandwidth is determined solely by the pulse's amplitude and not by its spectral composition.

The gapped graphene is a real system where the symmetry of graphene is relaxed by positioning on a incommensurate substrate.  For example, the bandgap in graphene placed on hexagonal boron nitride (h-BN) or silicon carbide (SiC) can be as large as $0.5~\mathrm{eV}$\cite{Conrad_et_al_PhysRevLett.115_2015_Gapped_Graphene_on_SiC,Ajayan_et_al_Review-JNN_2011_Band_Opening_in_Graphene}. On the other hand, it is a convenient generic model  
to study two-dimensional hexagonal-lattice semiconductors where the bandgap can be switched on gradually. In particular, as we show below in this Article, as the bandgap increases, so the valley polarization increases.

\section{MODEL AND MAIN EQUATIONS}
\label{Model_and_Equations}

\begin{figure}
\begin{center}\includegraphics[width=0.47\textwidth]{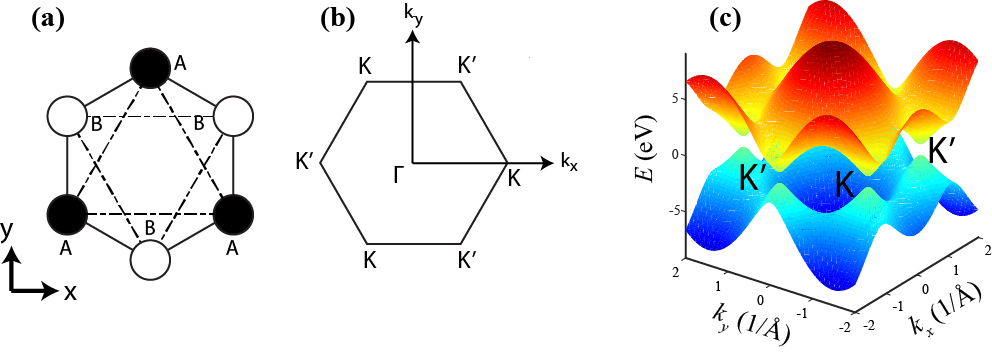}\end{center}
  \caption{(Color online) (a) Hexagonal lattice structure of graphene with sublattices A and B. (b) The first Brillouin zone of graphene with two valleys $K$ and $K^\prime$. (c) Energy dispersion as a function of crystal momentum for gapped graphene with the bandgap of 1 eV.}
  \label{fig:Energy}
\end{figure}

The unit cell of the honeycomb lattice of the gapped graphene model is shown in Fig.\ \ref{fig:Energy}(a) where the difference between the two sublattices, $A$ and $B$ breaks the $\mathcal P$ symmetry causing the bandgap to open up. The first Brillouin zone of the reciprocal lattice is shown in Fig.\ \ref{fig:Energy}(b) where the energy-dispersion symmetry between the $K$ and $K^\prime$ valleys is protected by the $\mathcal T$ symmetry.

We consider an ultrashort optical pulse with the duration of less than $5~\mathrm{fs}$. For such a short pulse, since  
the electron relaxation time is typically longer than $10~\mathrm{fs}$ for two dimensional materials\cite{Hwang_Das_Sarma_PRB_2008_Graphene_Relaxation_Time, Breusing_et_al_Ultrafast-nonequilibrium-carrier-dynamics_PRB_2011, theory_absorption_ultrafast_kinetics_graphene_PRB_2011, Ultrafast_collinear_scattering_graphene_nat_comm_2013, Gierz_Snapshots-non-equilibrium-Dirac_Nat-Material_2013, Nonequilibrium_dynamics_photoexcited_electrons_graphene_PRB_2013},
we can assume that the electron dynamics in the field of the pulse is coherent and can, therefore, be described by the time-dependent Schr\"odinger equation (TDSE), which has the following form
\begin{equation}
i\hbar \frac{{d\Psi }}{{dt}} = { H(t)} \Psi  
\label{Sch}
\end{equation}
with Hamiltonian in the space gauge as 
\begin{equation}
{ H}(t) = { H}_0 - e{\bf{F}}(t){\bf{r}},
\label{Ht}
\end{equation}  
where $\mathbf F(t)$ is the  pulse's electric field, $e$ is electron charge, and $H_0$ is the Hamiltonian of the solid in the absence of the optical field. We will approximate $H_0$ as a 
 nearest neighbor tight-binding Hamiltonian for gapped graphene \cite{Kjeld_et_al_PhysRevB.79.113406_2009_Gapped_Graphene_Optical_Response},
\begin{eqnarray}
H_0=\left( {\begin{array}{cc}
   \frac{\Delta_g}{2} & \gamma f(\mathbf k) \\
   \gamma f^\ast(\mathbf k) & -\frac{\Delta_g}{2} \\
  \end{array} } \right).
\label{H0}
\end{eqnarray}
Here $\Delta_g$ is the finite gap between the CB and the VB,
$\gamma= -3.03$ eV is hopping integral, and
\begin{equation}    
f(\mathbf k)=\exp\Big(i\frac{ak_y}{\sqrt{3}}\Big )+2\exp\Big(-i\frac{ak_y}{2\sqrt{3}}\Big )\cos{\Big(\frac{ak_x}{2}\Big )},
\end{equation}
where $a=2.46~\mathrm{\AA}$ is the lattice constant. The energies of CB and VB can be found from the above Hamiltonian, $H_0$, as follows
\begin{eqnarray}
E_{c}(\mathbf k)&=&+\sqrt{\gamma ^2\left |{f(\mathbf k)}\right |^2+\frac{\Delta_g ^2}{4} }~~,
\nonumber \\
E_{v}(\mathbf k)&=& -\sqrt{\gamma ^2\left |{f(\mathbf k)}\right |^2+\frac{\Delta_g ^2}{4}}~~,
\label{Energy}
\end{eqnarray}
where $c$ and $v$ stand for the CB and VB, respectively. The energy dispersion is shown in Fig. \ref{fig:Energy}(c).


In solids, an applied electric field generates both the intraband (adiabatic) and interband (non-adiabatic) electron dynamics. The intraband dynamics is determined by the Bloch acceleration theorem \cite{Bloch_Z_Phys_1929_Functions_Oscillations_in_Crystals} for time evolution of the crystal momentum, $\mathbf k$, dependence on time $t$
\begin{equation}
\hbar \frac{{d{\bf{k}}}}{{dt}} = e{\bf{F}}(t).
\label{acceleration}
\end{equation}
From this equation, for an electron with an initial crystal momentum ${\bf q}$, one finds 
\begin{equation}
{{\bf{k}}}({\bf{q}},t) = {\bf{q}} + \frac{e}{\hbar }\int_{ - \infty }^t {{\bf{F}}({t^\prime})d{t^\prime}}. 
\label{kvst}
\end{equation}

The corresponding adiabatic wave functions, which are the solutions of Schr\"odinger equation (\ref{Sch}) within a single band $\alpha$ without an interband coupling, are the well-known Houston functions \cite{Houston_PR_1940_Electron_Acceleration_in_Lattice},
\begin{equation}
 \Phi^\mathrm{(H)}_{\alpha {\bf q}}({\bf r},t)=\Psi^{(\alpha)}_{\bf{k}(\bf q,t)} ({\bf r})\exp\left(i\phi^{(\mathrm D)}_{\mathrm{\alpha}}({\bf q},t)+i\phi^{(\mathrm B)}_{\mathrm{\alpha}}({\bf q},t)\right),
\label{Houston}
\end{equation}
where $\alpha=v,c$ for the VB and CB, respectively, and $ \mathrm{\Psi^{(\alpha)}_{{\mathbf k}}} $ are the Bloch-band eigenfunctions in the absence of the pulse field. 
The dynamic phase, $\phi^\mathrm{(D)}_{\mathrm \alpha}$, and the geometric (Berry) phase, $\phi^\mathrm{(B)}_{\mathrm \alpha}$, are defined by the following expressions
\begin{eqnarray}
\phi^\mathrm{(D)}_{\alpha}(\mathbf q,t)= -\frac{1}{\hbar} \int_{-\infty}^t dt^\prime E_\mathrm \alpha[\mathbf k (\mathbf q,t^\prime)],
 \label{phi}
 \\ 
 \phi^\mathrm{(B)}_{\mathrm \alpha}(\mathbf q,t)= \frac{e}{\hbar} \int_{-\infty}^t dt^\prime \mathbf F(t^\prime) \mathbfcal{A}^{\mathrm{(\alpha \alpha)}}[\mathbf k (\mathbf q,t^\prime)].
 \label{phi}
\end{eqnarray}
Here  $\mathbfcal{A}^{(\alpha\alpha)}=\left\langle \Psi^{(\alpha)}_\mathbf q  |   i\frac{\partial}{\partial\mathbf q}|\Psi^{(\alpha)}_\mathbf q   \right\rangle $ is the (intraband) Berry connection for band $\alpha$. 

Note that the original Houston functions, as introduced in Ref.\ \onlinecite{Houston_PR_1940_Electron_Acceleration_in_Lattice}, were lacking the geometric phase because the periodic Bloch functions were considered real. As was introduced by Berry \cite{Berry_Phase_Proc_Royal_Soc_1984}, the phase of quantum mechanical wave functions is important as it defines the topology of electronic states. The phase accumulated along a closed adiabatic loop in a space of parameters, which is called geometric or Berry phase\cite{Berry_Phase_Proc_Royal_Soc_1984}, plays a fundamental role in many quantum mechanical and classical wave phenomena. 

In our case, the Berry phase appears due to the Bloch motion of an electron along a closed trajectory in the reciprocal (crystal momentum) space. It would anyway appear as a result of the numerical solution of Schr\"odinger equation. We have added this phase prior to the numerically solving the problem. Fundamentally,  this does not change the final result because both the original Houston functions and our modified functions of Eq.\ (\ref{Houston}) form complete basis sets. However, for our choice of the basis, coupling amplitude $\mathbfcal{\hat A}$ of Eq.\ (\ref{AA}) has only off-diagonal matrix elements. This means that we have exactly treated the intraband dynamics, and only the interband transitions need to be taken into account numerically. This simplifies analytical theory and makes the numerical calculations much more efficient. From the point of view of general theoretical classification, our theory is based on the interaction representation in an exact adiabatic basis.

In the present model, the Berry connection can be found analytically for the both VB and CB as
\begin{eqnarray}
\mathcal{A}_{x}^{(cc)}(\mathbf k)&=&\frac{-a\gamma ^2}{\gamma ^2 |f(\mathbf k)|^2+(\Delta_g/2-E_c)^2}
\nonumber \\
& \times &\sin \frac{\sqrt{3}ak_y}{2}\sin{\frac{ak_x}{2}},
 \label{Axcc}
 \\
\mathcal{A}_{y}^{(cc)}(\mathbf k)&=&\frac{a\gamma ^2}{\sqrt{3}(\gamma ^2 |f(\mathbf k)|^2+(\Delta_g/2-E_c)^2)}\nonumber\\
 &\times& \Big(\cos{ ak_x}-\cos{\frac{\sqrt{3}ak_y}{2}}\cos{\frac{ak_x}{2}}\Big),
 \label{Aycc}
 \\
\mathcal{A}_{x}^{(vv)}(\mathbf k)&=&\frac{-a\gamma ^2}{\gamma ^2 |f(\mathbf k)|^2+(\Delta_g/2+E_c)^2}
\nonumber \\
 &\times&\sin \frac{\sqrt{3}ak_y}{2}\sin{\frac{ak_x}{2}},
 \label{Axvv}
 \\
\mathcal{A}_{y}^{(vv)}(\mathbf k)&=&\frac{a\gamma ^2}{\sqrt{3}(\gamma ^2 |f(\mathbf k)|^2+(\Delta_g/2+E_c)^2)}\nonumber\\
 &\times &\Big(\cos{ ak_x}-\cos{\frac{\sqrt{3}ak_y}{2}}\cos{\frac{ak_x}{2}}\Big).
 \label{Ayvv}
\end{eqnarray}

The interband electron dynamics is determined by solutions of  TDSE (\ref{Sch}).  Such  solutions can be expanded in the adiabatic basis of the Houston functions, $\Phi^{(H)}_{\alpha {\bf q}}({\bf r},t)$,
\begin{equation}
\Psi_{\bf q} ({\bf r},t)=\sum_{\alpha=c,v}\beta_{\alpha{\bf q}}(t) \Phi^{(H)}_{\alpha {\bf q}}({\bf r},t),
\end{equation}
where 
$\beta_{\alpha{\bf q}}(t)$ are expansion coefficients.

These expansion coefficients satisfy the following system of differential equations 
\begin{equation}
i\hbar\frac{\partial B_\mathbf q(t)}{\partial t}= H^\prime(\mathbf q,t){B_\mathbf q}(t)~,
\label{Schrodinger}
\end{equation}
where wave function (vector of state) $B_q(t)$ and Hamiltonian $ H^\prime(\mathbf q,t)$ are defined as 
\begin{eqnarray}
B_\mathbf q(t)&=&\begin{bmatrix}\beta_{c\mathbf q}(t)\\ \beta_{v\mathbf q}(t)\\ \end{bmatrix}~,\\ 
H^\prime(\mathbf q,t)&=&-e\mathbf F(t)\mathbfcal{\hat A}(\mathbf q,t)~,\\
\mathbfcal{\hat A}(\mathbf q,t)&=&\begin{bmatrix}0&\mathbfcal D^{(cv)}(\mathbf q,t)\\
\mathbfcal D^{(cv)}(\mathbf q,t)^\ast&0
\label{AA}\\
\end{bmatrix}~.
\end{eqnarray}
Here
\begin{eqnarray}
\mathbfcal D^\mathrm{(cv)}(\mathbf q,t)&=&
\mathbfcal A^\mathrm{(cv)}[\mathbf k (\mathbf q,t)]\times\nonumber \\
&&\exp\left(i\phi^\mathrm{(D)}_\mathrm{cv}(\mathbf q,t)+i\phi^\mathrm{(B)}_\mathrm{cv}(\mathbf q,t)\right)~,~~~
 \label{Q}
\\
\phi^\mathrm{(D)}_\mathrm{cv}(\mathbf q,t)&=&\phi^\mathrm{(D)}_\mathrm{v}(\mathbf q,t)-\phi^\mathrm{(D)}_\mathrm{c}(\mathbf q,t)
 \label{phi}
 \\ 
 \phi^\mathrm{(B)}_\mathrm{cv}(\mathbf q,t)&=&\phi^\mathrm{(B)}_\mathrm{v}(\mathbf q,t)-\phi^\mathrm{(B)}_\mathrm{c}(\mathbf q,t)\\
{\mathbfcal{A}}^\mathrm{(cv)}({\mathbf q})&=&
\left\langle \Psi^\mathrm{(c)}_\mathbf q  |   i\frac{\partial}{\partial\mathbf q}|\Psi^\mathrm{(v)}_\mathbf q   \right\rangle~,~~~
\label{D}
\end{eqnarray} 
where ${\mathbfcal A}^{(cv)}(\mathbf q)$ is a matrix element of the well-known non-Abelian Berry connection \cite{Wiczek_Zee_PhysRevLett.52_1984_Nonabelian_Berry_Phase, Xiao_Niu_RevModPhys.82_2010_Berry_Phase_in_Electronic_Properties, Yang_Liu_PhysRevB.90_2014_Non-Abelian_Berry_Curvature_and_Nonlinear_Optics},  and $\phi^{\mathrm{(D)}}_{cv}(\mathbf q,t)$ and $\phi^{\mathrm{(B)}}_{cv}(\mathbf q,t)$ are the transitional dynamic phase and Berry (geometric) phase, respectively. Note that the interband dipole matrix element, which determines optical transitions between the VB and CB at a crystal momentum $\mathbf q$, is expressible in terms of the interband Berry connection as $\mathbf D^{(cv)}(\mathbf q)=e \mathbfcal{A}^{(cv)}(\mathbf q)$.

The non-Abelian Berry connection matrix elements can also be found analytically as
\begin{eqnarray}
&\mathcal{A}&_{x}^\mathrm{(cv)}(\mathbf k)=-\mathcal N\frac{a}{2|f(\mathbf k)|^2}\Bigg( \sin\frac{ak_x}{2}\sin\frac{a\sqrt{3}k_y}{2}
\nonumber\\
&&+i \frac{\Delta_g}{2E_c}\Big(\cos \frac{a\sqrt{3}k_y}{2}\sin \frac{ak_x}{2}+\sin{ak_x}\Big)\Bigg)~,~~~
 \label{Ax}
\\
&\mathcal{A}&_{y}^\mathrm{(cv)}(\mathbf k)=\mathcal N\frac{a}{2\sqrt{3}|f(\mathbf k)|^2}\Bigg( -1-\cos\frac{a\sqrt{3}k_y}{2}\cos\frac{ak_x}{2}
\nonumber\\
&&+2\cos ^2 \frac{ak_x}{2}-i \frac{3\Delta_g}2{E_c}\sin \frac{a\sqrt{3}k_y}{2}\cos \frac{ak_x}{2}\Bigg)~,~~~
\label{Ay}
\end{eqnarray}
where
\begin{equation}
\mathcal N=\frac{\left|\gamma f(\mathbf k)\right|}{\sqrt{\frac{\Delta_g ^2}{4}+\left|\gamma f(\mathbf k)\right|^2}}~.
\end{equation}
Schr\"odinger equation (\ref{Schrodinger}) completely describes the dynamics of the system. A formal general solution of 
this equation can be presented in terms of the evolution operator, $\hat S(\mathbf q,t)$, as follows
 \begin{eqnarray}
 B_\mathbf q (t)&=&\hat S(\mathbf q,t)B_\mathbf q (-\infty)~,
 \label{Bq} \\
 \hat S(\mathbf q,t)&=&\hat T \exp\left[i\int_{-\infty}^t \mathbfcal{\hat A}(\mathbf q,t^\prime)d\mathbf k(\mathbf q, t^\prime)\right]~,
 \label{S}
 \end{eqnarray}
where $\hat T$ is the well known time ordering operator \cite{Abrikosov_Gorkov_Dzialoshinskii_1975_Methods_of_Quantum_Field_Theory},  the 
integral is affected along the Bloch trajectory [Eq.\ (\ref{kvst})], and $d\mathbf k(\mathbf q,t)=\frac{e}{\hbar}\mathbf F(t)dt$.

After the pulse, the electron returns to its initial position in the reciprocal space, $\mathbf k(\mathbf q,\infty)=\mathbf q$, i.e., the Bloch trajectory is necessarily closed. The  evolution operator after the pulse, correspondingly, becomes
\begin{equation}
\hat S(\mathbf q,\infty)=\hat T \exp\left[i\oint \mathbfcal{\hat A}(\mathbf q,t)\,d\mathbf k(\mathbf q, t)\right]~.
\label{Sinfty}
\end{equation}

\section{Results and Discussion}
\label{Results}

We apply an ultrafast single-oscillation chiral (circularly-polarized) optical pulse incident normally to the gapped graphene layer. The electric field,  $\mathbf F$=($F_x$, $F_y$), 
of this pulse is set to have the following form 
\begin{eqnarray}
F_x&=&F_0(1-2u^2)e^{-u^2}
\label{Fx}
\\
F_y&=&\pm 2F_0ue^{-u^2}~,
\label{Fy}
\end{eqnarray}
where $F_0$ is the amplitude of the pulse, $u=t/\tau$, and $\tau$ is a characteristic time of the optical oscillation (below we choose $\tau$ = 1 fs). Here 
$\pm$ determines the handedness of the chiral pulse, where the ``+" sign corresponds to the right-handed circular polarization, and the ``-" sign corresponds to the left-handed circular polarization. Such an optical pulse contains a broad frequency spectrum with the mean frequency of $\omega=1.4~\mathrm{eV/\hbar}$ and bandwidth $\sim 2$ eV.
 
For a given waveform of the pulse, we solve TDSE with the following initial condition: $\beta_{\mathrm c\mathbf{q}}(-\infty )=0$, $\beta_{\mathrm v\mathbf{q}}(-\infty)=1$, which means that initially the conduction band is completely empty, and the valence band is fully occupied. The applied pulse transfers electrons from the VB to the CB, resulting in a nonzero residual (after the pulse) CB population $N_\mathrm{CB}^\mathrm{(res)}({\mathbf q})=|\beta_{\mathrm c\mathbf{q}}(\infty)|^2$~.

\begin{figure}
\begin{center}\includegraphics[width=0.47\textwidth]{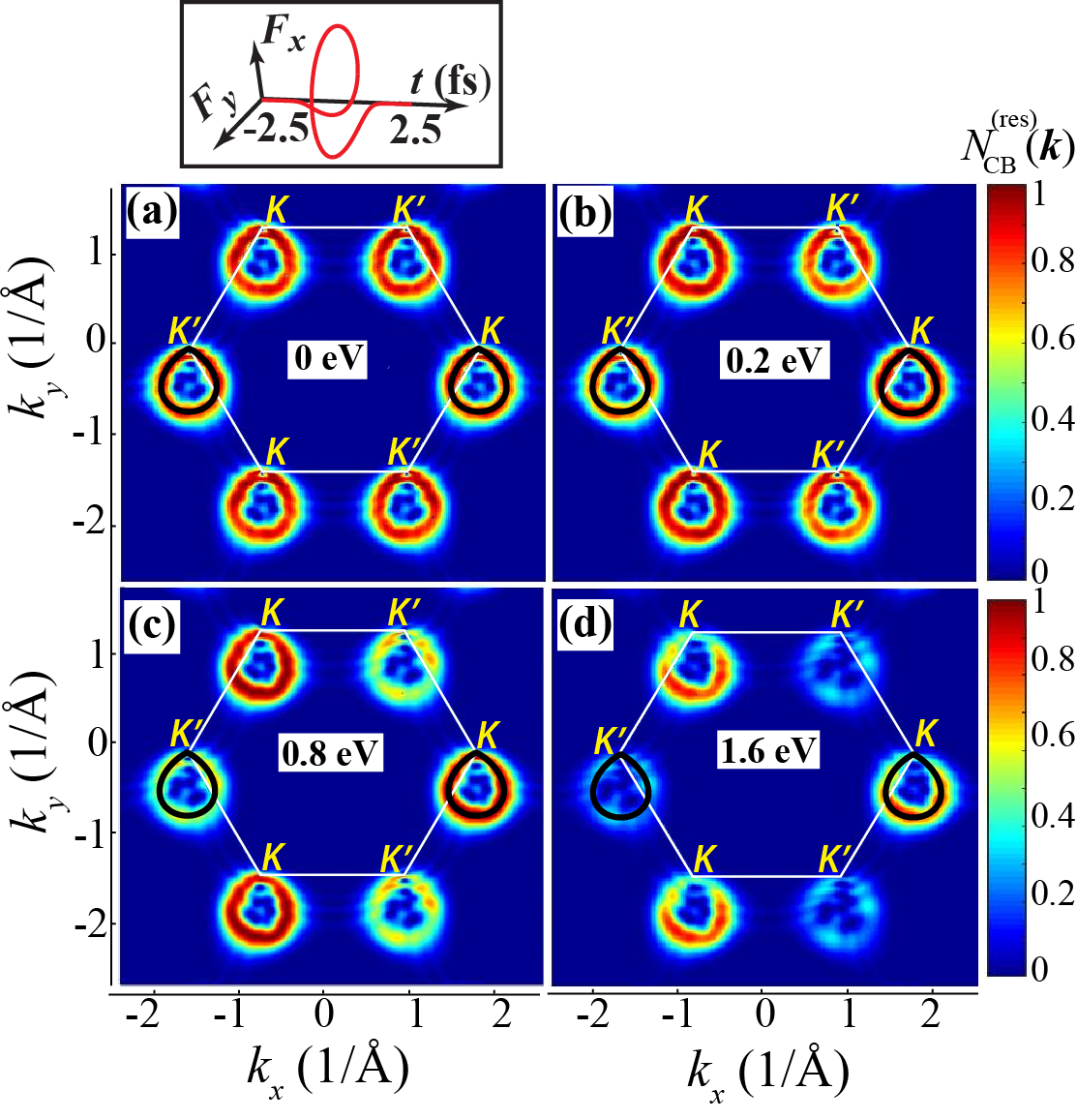}\end{center}
  \caption{(Color online) Residual CB population $N\mathrm{^{(res)}_\mathrm{CB}}(\mathbf{k})$ for gapped graphene in the extended zone picture. The applied optical pulse has right  circular polarization with the amplitude of $F_0=0.5 ~\mathrm{V\AA^{-1}}$. Inset: Waveform of the pulse $\mathbf F(t)=\{F_x(t), F_y(t)\}$ as a function of time $t$.  The white solid line shows the boundary of the first Brillouin zone with the $K, K^\prime$-points indicated.The bandgap is 0 (a), 0.2 eV (b), 0.8 eV (c), and 1.6 eV (D), as indicated on the corresponding panels. The separatrix is shown by a solid black line.
  }
  \label{fig:1RC_F0=0p5VpA_gapped_graphene}
\end{figure}%

\begin{figure}
\begin{center}\includegraphics[width=0.47\textwidth]{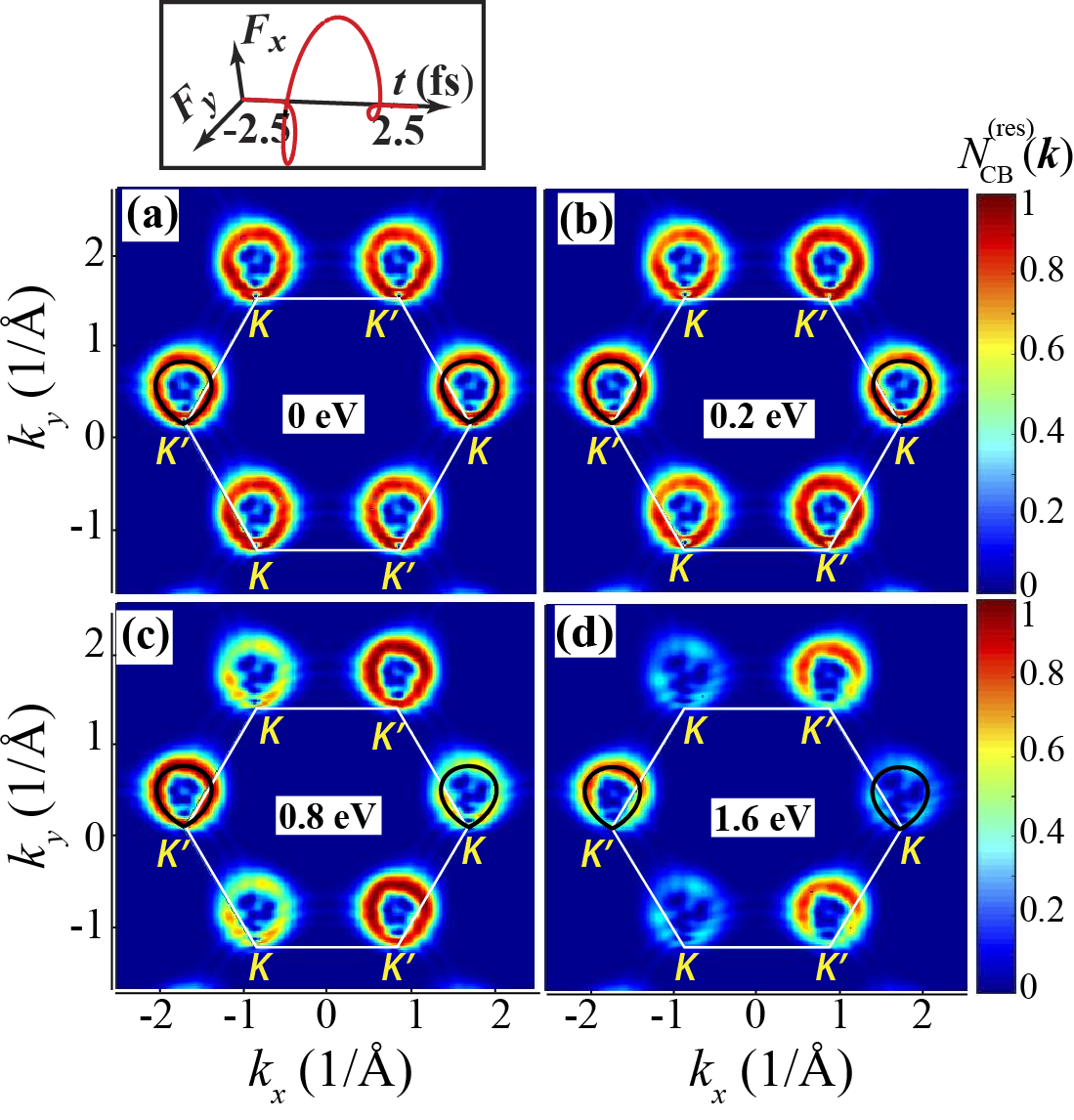}\end{center}
\caption{(Color online) Same as in Fig.\ \ref{fig:1RC_F0=0p5VpA_gapped_graphene} but for the pulse with left-hand circular polarization.}
 \label{fig:1CT_F0=0p5VpA_gapped_graphene}
\end{figure}

The residual CB population distribution in the reciprocal space is shown for gapped graphene   in Fig.\ \ref{fig:1RC_F0=0p5VpA_gapped_graphene}
for various bandgaps $\Delta_g$. The excitation pulse has right-handed circular polarization and the amplitude of $0.5~\mathrm{V\AA^{-1}}$. 

For the case of graphene, $\Delta_g=0$, Fig.\ \ref{fig:1RC_F0=0p5VpA_gapped_graphene}(a), for both $\mathrm K$ and $\mathrm K^\prime$ valleys, one can see a bright caustic of the electron population only along closed lines whose apexes are at the $\mathrm K$ and $\mathrm K^\prime$ points. Such a closed line has been introduced earlier and called a separatrix \cite{Stockman_et_al_PhysRevB.98_2018_Rapid_Communication_Topological_Resonances}. The separatrix is defined as a set of points (line) such so an electron Bloch trajectory originating from any point on this line passes exactly through the $\mathrm K$ (or $\mathrm K^\prime$) point during the pulse. Its analytical expression in a parametric form is \cite{Stockman_et_al_PhysRevB.98_2018_Rapid_Communication_Topological_Resonances}
\begin{equation}
{{\bf{k}}}(t) = {\bf{K}} - \frac{e}{\hbar }\int_{ - \infty }^t {{\bf{F}}({t^\prime})d{t^\prime}}~,
\label{separatrix}
\end{equation}
where $t$ is a parameter, and $\mathbf K$ is a crystal momentum for the corresponding $\mathrm K$- or $\mathrm K^\prime$-point. 

Only the electrons with the initial crystal momenta close to the separatrix will pass close to the regions with high interband coupling, which are the vicinities of the corresponding $K$- or $K^\prime$ points and, consequently, have a large probability to undergo transitions $\mathrm{VB\to CB}$. Therefore the separatrix will be surrounded by a bright closed arc of a high electron population -- a caustic, as clearly seen in Fig.\ \ref{fig:1RC_F0=0p5VpA_gapped_graphene}(a). In this case, the electron distribution at the $\mathrm K$-points vs.\ that at the $\mathrm K^\prime$-points is almost (though not perfectly) identical, which tells one that only a very small chirality in the electron distributions is induced.

For $\Delta_g =0$, the CB population distribution is symmetric with respect to the $y$-axis, and total population of both $\mathrm K$ and $\mathrm K^\prime$ valleys are equal, resulting in a zero total valley polarization. This can be explained by the presence of both the $\mathcal T$ symmetry and mirror-reflection symmetry with respect to both the $xz$-plane ($\mathcal P_{xz}$-symmetry) and the $yz$-pane ($\mathcal P_{yz}$-symmetry). In fact, the application of the $\mathcal T$ symmetry operation leads to change $\mathbf k\to -\mathbf k$, while $\mathcal P_{xz}$ operation implies that $k_y\to -k_y$. Thus application of the combined $\mathcal TP_{xz}$ symmetry operation leads to the change $k_x\to -k_x$ implying symmetry of the electron distribution with respect to the mirror reflection in the $yz$ plane, which is clearly seen in Fig.\ \ref{fig:1RC_F0=0p5VpA_gapped_graphene}(a).

Breaking the $\mathcal P_{xz}$ symmetry of graphene by introducing an asymmetry between sublattices A and B causes opening of a bandgap, $\Delta_g $. It also eliminates the exact symmetry of the electron distributions with respect to $\mathcal P_{yz}$ reflection discussed above in the previous paragraph  and, consequently, will cause a valley polarization. The calculated CB population distributions for $\Delta_g>0 $ are shown in Figs.\ \ref{fig:1RC_F0=0p5VpA_gapped_graphene}(b)-(d). With increasing the bandgap, the CB population at the $\mathrm K^\prime$ valley is strongly suppressed, resulting in a correspondingly increasing valley polarization. For a ``resonant'' bandgap, i.e., $\Delta_g\sim \hbar/(2\tau)$, the valley polarization is almost perfect -- see Fig.\ \ref{fig:1RC_F0=0p5VpA_gapped_graphene}(d).

Note that the phenomena that we consider are non-resonant in the common sense: the pulse is too short and, correspondingly, too broad spectrally to cause the conventional frequency-defined resonances. The optimum duration of the pulse that causes the maximum valley polarization is defined mostly by the bandgap: $\tau\sim \hbar/(2\Delta_g)$. The dependence of the valley polarization on the pulse duration is very smooth. However, one must point out that the bandgap, $\Delta_g$. should not be too small, so the pulse duration is still short enough with respect to the electron collision times, $\gtrsim 10$ f. Thus the carrier  frequency of the pulse should be from mid-ir and into the visible range.

For the left-hand circularly-polarized pulse, the obtained electron distributions are shown in Fig.  \ref{fig:1CT_F0=0p5VpA_gapped_graphene}. These data are directly calculated but could also be obtained from those in  Fig.\ \ref{fig:1RC_F0=0p5VpA_gapped_graphene} by the application of the $\mathcal T$ reversal. Under such a transformation, the chirality of the pulse is changed to the opposite, and the crystal momentum is inverted, $\mathbf k\to-\mathbf k$. Consequently, the electron CB-population distributions in Fig.\ \ref{fig:1CT_F0=0p5VpA_gapped_graphene} are obtained from those in Fig.\ \ref{fig:1RC_F0=0p5VpA_gapped_graphene} by the central inversion, i.e., transformation $\mathcal P_\mathrm O=\mathcal P_{xz}\mathcal P_{yz}$.

 
\section{Topological Resonances in Gapped Graphene}
\label{TRGG}

We attribute the valley polarization of gapped graphene in the field of a chiral pulse to be due to the recently introduced effect of topological resonance \cite{Stockman_et_al_PhysRevB.98_2018_Rapid_Communication_Topological_Resonances}. The topological resonance occurs because of the interference and compensation of the two phases: the dynamic phase and the topological phase, which is a combination of the geometric phase and the phase of the dipole matrix element,
\begin{equation}
 \phi_{c v}^\mathrm{(tot)}(\mathbf q,t)=\phi^\mathrm{(B)}_\mathrm{cv}(\mathbf q,t)+\phi_{c \mathrm v}^{(\mathrm A)}(\mathbf q,t)+\phi^\mathrm{(D)}_\mathrm{c \mathrm v}(\mathbf q,t)~, 
 \label{phi_tot}
 \end{equation} 
where the dipole matrix element phase, $\phi_{ \mathrm {cv}}^{(\mathrm A)}(\mathbf q,t)$, is defined as 
\begin{equation}
\phi_{ \mathrm {cv}}^{(\mathrm A)}(\mathbf q,t)=\arg\left\{ \mathbfcal{A}^{ \mathrm {(cv)}}[\mathbf k(\mathbf q,t)]\mathbf n(t) \right\}~,
\label{phiT}
\end{equation}
and $\mathbf n(t)=\mathbf F(t)/F(t)$ is a unit vector tangent to the Bloch trajectory.

\subsection{Topological Resonance in Perturbation Theory}
\label{PT}

To elucidate the origin of the topological resonances, we will consider a simplified theory for the case where the perturbation theory is applicable. In this case, Eq.\ (\ref{Sinfty}) becomes
\begin{equation}
\hat S(\mathbf q,\infty)=1+i\oint \mathbfcal{\hat A}(\mathbf q,t)\,d\mathbf k(\mathbf q, t)~.
\label{SPT}
\end{equation}
From this, using Eq.\ (\ref{Bq}), we find the residual CB population in the first order of perturbation theory,
\begin{equation}
 n_\mathrm c=\left\vert \oint \mathbfcal{ A}^{ \mathrm {(cv)}}[\mathbf k(\mathbf q,t)]d\mathbf k(\mathbf q,t)\right\vert^2~.
 \label{n_alpha}
 \end{equation}
We rewrite Eq.\ (\ref{n_alpha}) in terms of the phase of effective interband coupling  $\mathbfcal{\hat A}^{ \mathrm {(cv)}}$ as
 \begin{eqnarray}
 n_\mathrm{c}& =& \bigg\vert\oint\Big\vert \mathbfcal{ A}^{\mathrm{(cv)} }[\mathbf k(\mathbf q,t)]\mathbf n(t) \Big\vert\times 
 \nonumber \\
& & \exp\left[i\phi_{\mathrm{cv} }^\mathrm{(tot)}(\mathbf q,t)\right]d k(\mathbf q,t)\bigg\vert^2~,
 \label{n_alpha_phases}
 \end{eqnarray}
 where $\mathbf n(t)=\mathbf F(t)/F(t)$ is a unit vector tangent to the Bloch trajectory;  
 the total phase, $\phi_{c v}^\mathrm{(tot)}(\mathbf q,t)$, is given above by Eq.\ (\ref{phi_tot}).
 
\begin{figure}
\begin{center}\includegraphics[width=0.47\textwidth]{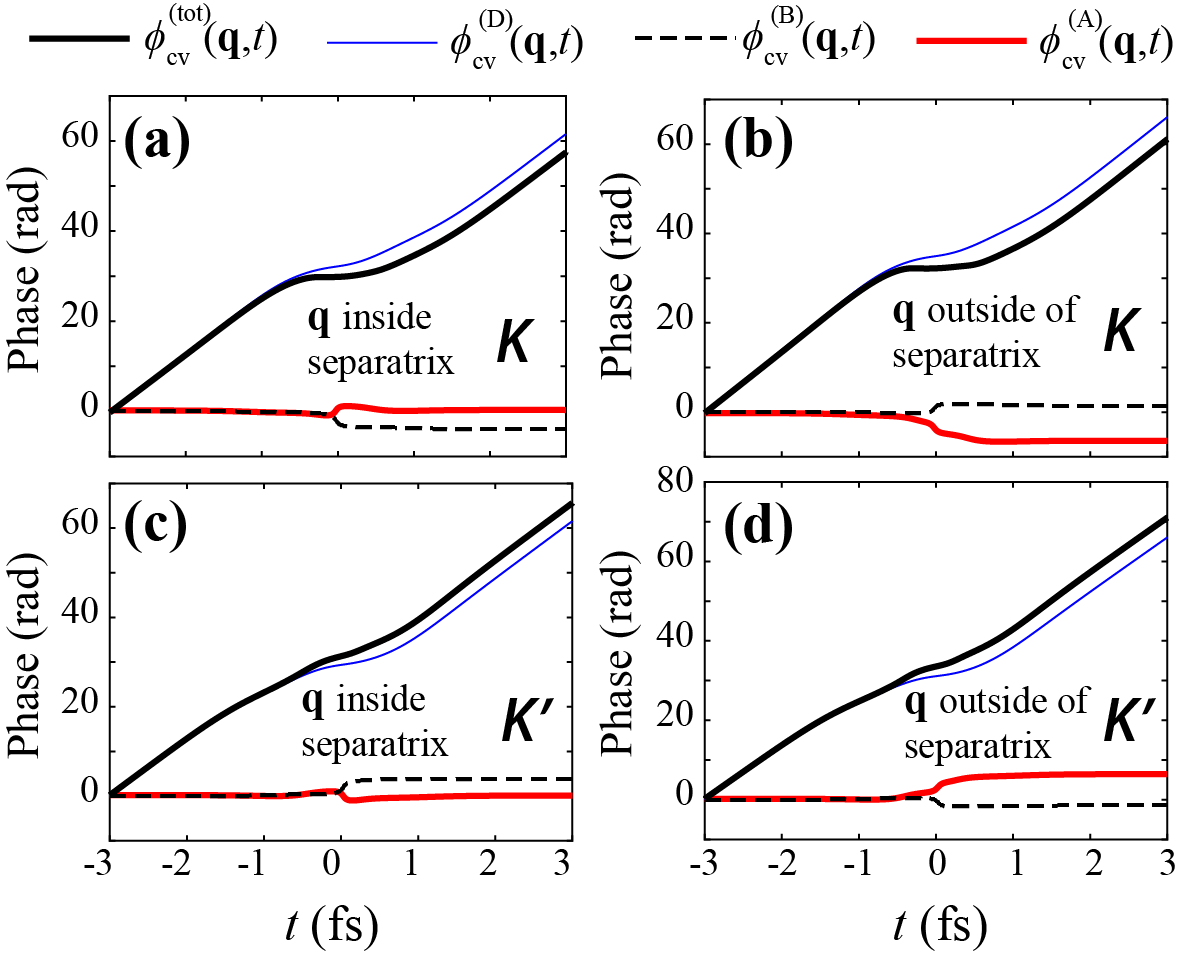}\end{center}
\caption{(Color online) 
For a pulse with right-hand circular polarization and amplitude $F_0=0.5 ~\mathrm{V\AA^{-1}}$, phases $\phi_\mathrm{cv}^\mathrm{(tot)}(\mathbf q,t)$, $\phi_\mathrm{cv}^\mathrm{(D)}(\mathbf q,t)$, $\phi_\mathrm{cv}^\mathrm{(B)}(\mathbf q,t)$, and $\phi_\mathrm{cv}^\mathrm{(T)}(\mathbf q,t)$ are shown for different initial wave vectors. (a) The initial wave vector,  $\mathbf q$, is inside the separatrix in the $K$ valley. (b) The initial wave vector, $\mathbf q$, is outside of the separatrix in the $K$ valley. (c) The initial wave vector,  $\mathbf q$, is inside the separatrix at the $K^\prime$ valley.  (D) The initial wave vector, $\mathbf q$, is outside of the separatrix in the $K^\prime$ valley. The separatrix is shown in Fig.\ \ref{fig:1RC_F0=0p5VpA_gapped_graphene} by a solid black line.
} 
\label{fig:phase}
\end{figure}


The topological resonance, which manifests itself as a caustic in the reciprocal space (i.e., as an arc  where the residual CB population is large), occurs when the total phase $\phi_\mathrm{cv}^\mathrm{(tot)}(\mathbf q,t)$ is stationary, i.e., the time variation of the topological phase cancels the time variation of the dynamic phase. The corresponding phases, $\phi_\mathrm{cv}^\mathrm{(tot)}(\mathbf q,t)$, $\phi_\mathrm{cv}^\mathrm{(D)}(\mathbf q,t)$, $\phi_\mathrm{cv}^\mathrm{(B)}(\mathbf q,t)$, and $\phi_\mathrm{cv}^\mathrm{(A)}(\mathbf q,t)$, are shown in Fig.\ \ref{fig:phase} for gapped graphene with the bandgap of $1.6~\mathrm{eV}$ for the right-hand circular polarization. We show these phases for different values of initial wave vector (crystal momentum), $\mathbf q$, which are in the vicinity of the separatrix either inside or outside of it.  Since the magnitude of the interband coupling is strongest near the $K$- and $K^\prime$ points, which corresponds to $t=0$ in Fig.\ \ref{fig:phase}, we need to analyze the behavior of the total phase at $t$ close to zero. 

To start, we notice that while phases $\phi_\mathrm{cv}^\mathrm{(B)}(\mathbf q,t)$ and $\phi_\mathrm{cv}^\mathrm{(A)}(\mathbf q,t)$, for a given valley, have opposite signs for points $\mathbf q$ inside and outside of the separatrix, their sum [the total topological phase, $\phi_\mathrm{cv}^\mathrm{(tot)}(\mathbf q,t)$] has the same behavior for points close to the separatrix, both inside and outside of it. Thus, the time dependence of the total topological phase strongly depends on the valley, $K$ or $K^\prime$, and not very significantly on the position of $\mathbf q$, situated near the separatrix either inside or outside of it. As we see in the case of Fig.\ \ref{fig:phase}, the total phase, $\phi_\mathrm{cv}^\mathrm{(tot)}(\mathbf q,t)$, is stationary at $t=0$ only for the $K$-valley [Figs.\ \ref{fig:phase} (a) and (b)] but not for the $K^\prime$-valley [Figs.\ \ref{fig:phase} (c) and (d)]. Thus, for the right-hand circular-polarized pulse, the  topological resonance favors the residual population of the $K$-valley while  the residual CB population of the $K^\prime $-valley is relatively small --  cf. Fig.\ \ref{fig:1RC_F0=0p5VpA_gapped_graphene}. 

In contrast, for the left-hand circularly-polarized pulse, the topological phase for a given valley has the opposite sign. Consequently, it is the $K^\prime$ valley that is predominantly populated, and $K$-valley is almost unpopulated, i.e., the valley polarization is exactly opposite -- cf. Fig.\ \ref{fig:1CT_F0=0p5VpA_gapped_graphene}.


\begin{figure}
\begin{center}\includegraphics[width=0.47\textwidth]{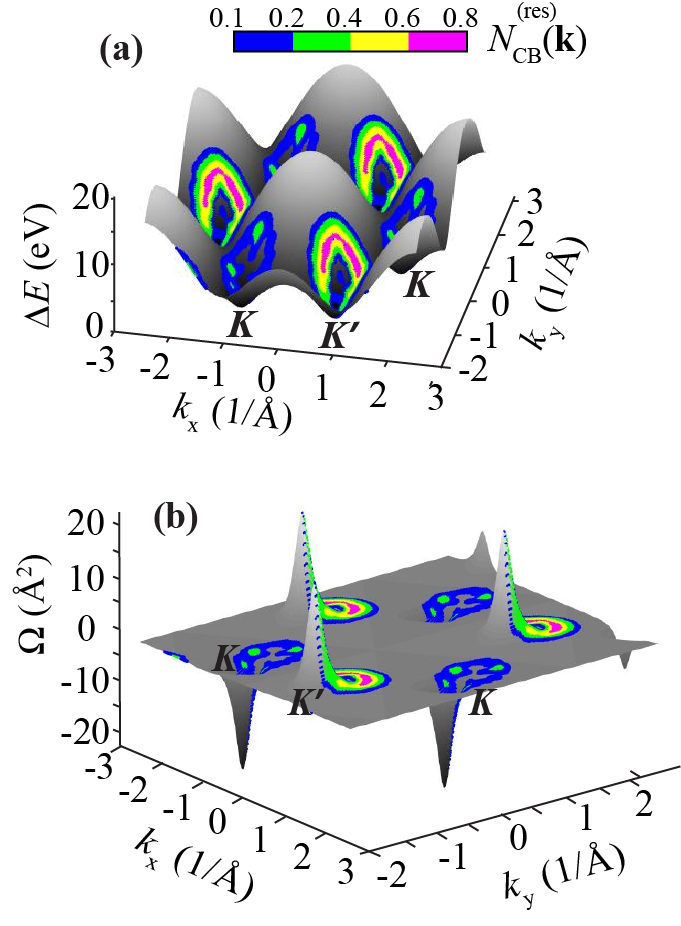}\end{center}
 \caption{(Color online) Residual CB population $N_\mathrm{CB}^\mathrm{(res)}(\mathbf k)$ of gapped graphene with bandgap $\Delta_g=2$ eV is shown with color-coding on the background of (a) the corresponding  excitation energy and (b) the corresponding Berry curvature in the extended zone picture. The applied pulse is left-handed with the amplitude of 0.5 $\mathrm{V/\AA}$ }.
\label{fig:CB_Berry_Energy}
\end{figure}

\subsection{Energy Band and Valley Polarization for Topological Resonances}
\label{EBTR}

In one respect, the topological resonance is similar to a regular resonance, which occurs when the frequency, $\hbar\omega$, of a long pulse coincides with the bandgap
between the CB and VB states. For both resonances, the dynamic phase, $\phi_{c \mathrm v}^{(\mathrm D)}(\mathbf q,t)$, is canceled by another phase, which is the total topological phase, $\phi_{c \mathrm v}^{(\mathrm A)}(\mathbf q,t)+\phi^\mathrm{(B)}_\mathrm{c \mathrm v}(\mathbf q,t)$, for the topological resonance and the phase of the pulse, $\phi _ p = \omega t$, for the regular resonance. 

However, there is a fundamental difference between these two resonances. Namely, the 
regular resonance depends on the frequency of the pulse, but not on its amplitude, while the topological resonance depends mostly on the amplitude of the pulse, but not on its 
frequency. The regular resonance would be represented by a CB population close to the bottom of the corresponding valley ($K$- or $K^\prime$-points) where pulse frequency almost exactly bridges the bandgap, $\hbar\omega\approx\Delta_g$. In a sharp contrast, for the topological resonance the electrons promoted to the CB occupy the states along the separatrix in a very wide energy band of $\sim 12$ eV -- see the next paragraph. This energy range increases with the pulse amplitude but does not depend on the pulse carrier frequency.

To illustrate these fundamental properties, we show in Fig.\ \ref{fig:CB_Berry_Energy}(a) the CB population distribution, $N_\mathrm{CB}^\mathrm{(res)}(\mathbf k)$, together with the corresponding excitation energy, $E_c(\mathbf k)-E_v(\mathbf k)$. These data show that the CB population is spread along the separatrix almost uniformly over an enormously large excitation-energy range of  $\sim 12$ eV. At the same time, our excitation pulse has the central frequency of 1.4 eV and the duration of $\sim 2$ fs, i.e., its bandwidth is $\lesssim 2.5$ eV. Thus the regular resonance cannot be responsible for such a large CB population, $N_\mathrm{CB}^\mathrm{(res)}(\mathbf k)\sim 1$, which is almost uniformly spread over the $\sim 12$-eV bandwidth along the separatrix in one type of the valley ($K$ or $K^\prime$) selected by the chirality of the pulse. This is a hallmark of the topological resonance. Additionally, Fig.\ \ref{fig:CB_Berry_Energy}(b) indicates that the region of the excited electrons in the reciprocal space always includes the center of the Berry curvature, i.e., the $K$- or $K^\prime$-point, which is an indication of its origin: in a sense, this region is an image of the region of the high Berry curvature, which is produced by the Bloch motion of the electrons. 

Finally, the number of electronic states available and, correspondingly, the maximum number of electrons that can be transferred to the CB by the regular resonance is limited by its small energy band due to the Pauli blocking. In a sharp contrast, this number for the topological resonance is orders of magnitude larger proportionally to its gigantic energy range. This is of a paramount importance for the application of the valley-polarized materials, in particular, for anomalous Hall effect.

\begin{figure}
\begin{center}\includegraphics[width=0.47\textwidth]{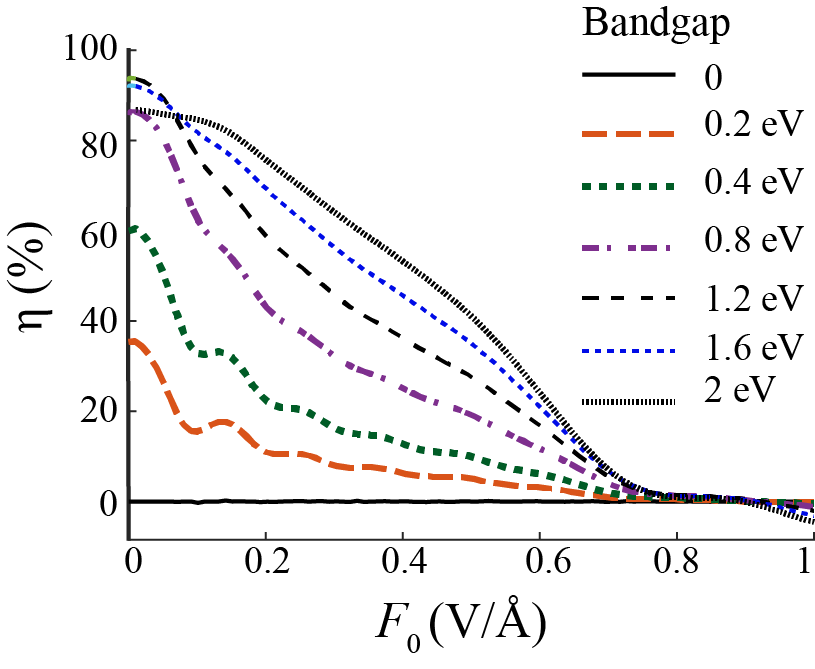}\end{center}
 \caption{(Color online) Valley polarization of gapped graphene as a function of the amplitude $F_0$ of the excitation right-handed pulse for different bandgaps, $\Delta_g$, 0, $0.2~\mathrm{eV}$, $0.4~\mathrm{eV}$, $0.8~\mathrm{eV}$, $1.2~\mathrm{eV}$, $1.6~\mathrm{eV}$, $2~\mathrm{eV}$. }
\label{fig:Valley_Polarization}
\end{figure}

For a given residual CB distribution, we define the valley polarization as
\begin{equation}
\eta=\frac{n_K-n_{K^\prime}}{n_K+n_{K^\prime}},
\end{equation}
where $n_K=\Sigma_{{\bf k} \in K} N^\mathrm{(res)}_{\mathrm{CB}}({\bf k})$ is the total CB population of the $K$ valley, i.e., the sum is over all $\bf k$ points at the $K$ valley, while 
$n_{K^\prime}$ is the same for the ${K^\prime}$ valley.

The valley polarization is shown in Fig.\ \ref{fig:Valley_Polarization} as a function of the pulse amplitude. As expected, the valley polarization is large and increases with 
the bandgap. For example, for a field amplitude of  $0.2~\mathrm{V/\AA}$, large valley polarization of more than $50$ percent is achieved in gapped graphene with the bandgap of 0.8 eV or higher. Even for bandgap as low as $0.4~\mathrm{eV}$, which can be experimentally achieved for graphene epitaxially grown on SiC substrate\cite{Conrad_et_al_PhysRevLett.115_2015_Gapped_Graphene_on_SiC}, the valley polarization is about $25$ percent (see Fig.\ \ref{fig:Valley_Polarization}). With increasing the field amplitude, the valley polarization decreases and even reverses its sign. The reason for this behavior is that for large field amplitudes, the electron trajectory, which starts at one valley, reaches the other valley during the pulse, which results in populating both of the two valleys. 

\subsection{Topological Resonances in Strong Field Case}
\label{TRSFC}

Above in Sec.\ \ref{PT} we have elucidated an origin of topological resonances in perturbation theory. Here we briefly consider a strong field case without invoking an assumption of perturbation theory. We will show that the condition of the topological resonances remains the same -- mutual compensation of the dynamic and topological phases as given by Eq.\ (\ref{phi_tot}).

Let us consider general properties of the evolution operator ($S$-matrix), $\hat S=\hat S(\mathbf q,\infty)$. Doing so, we follow the standard theory of the $S$-matrix \cite{Berestetskii_Lifshits_Pitaevskii_Quantum_Electrodynamics}. 

We introduce the transition amplitude operator, $\hat A$, by an expression
\begin{equation}
\hat S=1+i\hat A~.
\label{A}
\end{equation}
Then we take into account that the $S$-matrix is necessary unitary, i.e., 
\begin{equation}
\hat S^\dag \hat S=1~.
\label{SdagS}
\end{equation}
Substituting Eq.\ (\ref{A}), we obtain
\begin{equation}
A^\dag A =i\left(A^\dag-A\right)~.
\end{equation}
Averaging this equation over the initial state of the system, $\left|0\right\rangle$, we obtain the well-known optical theorem
\begin{equation}
\sum_f \left|\left\langle f\left|\hat A\right|0\right\rangle\right|^2=2\mathrm{Im}\left\langle 0 \left|\hat A\right|0\right\rangle~.
\label{OT}
\end{equation}
where the summation is extended over all final states $| f \rangle $. Taking into account that the left-hand side of this equation is the total population of all conduction bands, $n_{\mathrm c}^\mathrm{(tot)}$, and taking into account Eq.\ (\ref{A}) we arrive at the desired exact relation
\begin{equation}
n_{\mathrm c}^\mathrm{(tot)}=-2\mathrm{Re}\left\langle 0 \left|\hat S-1\right|0\right\rangle~.
\label{nctot}
\end{equation}

The $S$-matrix operator of Eq.\ (\ref{Sinfty}) contains in the exponent the operator of non-Abelian Berry connection, $\mathbfcal{\hat A}(\mathbf q,t)$ of Eq.\ (\ref{AA}), which possesses only the non-diagonal matrix elements. This implies that in the quantum mechanical average in the right-hand side of Eq.\ (\ref{nctot}) only even powers of  $\mathbfcal{\hat A}$ will contribute. Consequently, imaginary unity $i$ in the exponent of the $S$-matrix operator of Eq.\ (\ref{Sinfty}) will be eliminated. Thus the compensation of the dynamic and topological phases will cause a coherent accumulation of real contributions into $\left\langle 0 \left|\hat S-1\right|0\right\rangle$ and, thus, maximize the total population of the conduction bands,  $n_{\mathrm c}^\mathrm{(tot)}$, in Eq.\ (\ref{nctot}). This is precisely what we find solving numerically the Schr\"odinger equation for strong fields. This demonstrates that the phase-compensation condition of Eq.\ (\ref{phi_tot}) is the necessary and sufficient condition of the topological resonance irrespectively of the field strength.

\section{Conclusions}

We predict that large, ultrafast, valley-selective CB population can be induced in gapped graphene by a circularly-polarized single-oscillation optical pulse. The corresponding residual (after the pulse) valley polarization strongly depends on the bandgap. 
While the valley polarization is zero for the 
native gappless graphene, it monotonically increases with the bandgap, resulting in considerable valley polarization at the bandgap of $\Delta_g\approx 2$ eV. This bandgap value is defined by the length of the pulse (i.e., optical period), $\Delta_g\sim \hbar/(2\tau)$.

The origin of such valley-selective CB population is the topological resonance, which occurs because the electron wave functions gradually accumulate the topological (geometric) phases along the electron Bloch trajectory in the field of the pulse. 
Such gradual accumulation of the topological phase is possible only in gapped graphene where the corresponding Berry curvature (topological magnetic field) is extended over finite regions near the $K$ and $K^\prime$ points. In a sharp contrast, in gappless graphene, the Berry curvature is singularly concentrated at the 
corresponding Dirac points. The topological phases are opposite in the valleys of different types due to the time-reversal symmetry.



The accumulated topological phase interferes with the dynamic phase, which is related to the bandgap energy. This interference results in the topological resonance effect when the topological phase compensates the dynamic phase. Such compensation results in the total phase of the interband transition amplitude becoming almost stationary (time-independent). In such a case, the interband transition amplitude accumulates coherently in time leading to a large CB population, which is the topological resonance.  

The Bloch trajectories that cause the maximum population are those that pass through the regions of a high Berry curvature, which are in the vicinities of the $K$- and $K^\prime$-points. The initial crystal momenta of such trajectories lie in the vicinity of the curves that we called the separatrices. These are seen in the electron momentum distribution in the CB as caustics -- the bands of a high electron population along the separatrices. These caustics are clearly seen in the electron CB distributions presented in Figs.\ \ref{fig:1RC_F0=0p5VpA_gapped_graphene}, \ref{fig:1CT_F0=0p5VpA_gapped_graphene}, and \ref{fig:CB_Berry_Energy}.

Depending on the handedness of the circularly-polarized pulse, the topological resonance occurs predominantly in either $K$- or $K^\prime$-valleys, which induces a significant valley polarization in the gapped graphene. Experimentally, the distribution of the residual CB electron population and the corresponding valley polarization in the reciprocal space can be directly  observed by time resolved angle-resolved photoelectron emission spectroscopy (TR-ARPES) \cite{Freericks_et_al_Annal_Phys_2017_Superconductors_TR-ARPES_Theory, Chiang_et_al_PhysRevLett.107_Berry_Phase_in_Graphene_ARPES}.

The topological resonance is a general nonlinear phenomenon. It is due to  the electron orbital dynamics and is mostly independent from the electron spin. It occurs in any semiconductor or insulator material with a non-zero Berry curvature and invokes the Bloch states in the vicinity of the bandgap. In particular, its is pronounced for graphene-like two-dimensional materials possessing a hexagonal lattice with broken inversion symmetry. Such materials have the bandgap at the $K$- or $K^\prime$-points. In contrast,  in the III-V semiconductors such as GaAs and GaN,  the bandgap is at the $\Gamma$-point, which is a  time-reversal-invariant point implying a zero Berry curvature.  

The topological resonance only weakly depends on the carrier frequency but has a strong dependence on the amplitude of the pulse. This is in a sharp contrast to the regular resonance, which is  determined by the frequency of the radiation but not its amplitude. The electrons that populate the CB due to the topological resonance are distributed in an ultrawide energy range. For instance, for a pulse with the amplitude of $\sim 0.5$ V/\AA, this range is $\sim 12$ eV, which is much greater than the energy bandwidth of the pulse, $\hbar/(2\tau)\sim 2$ eV.
This is in a sharp contrast to the regular resonance where the excited electrons are distributes in a narrow energy region near the bandgap. This ultrawide energy width of the topological resonance also provides much more CB states available to obtain a high CB population and allows for ultrafast excitation of it within a fundamentally shortest time period -- just a single optical oscillation.

The phenomena associated with the topological resonance can be experimentally studied with a variety of approaches. As we have already mentioned above in this Section,  the resulting electron distributions can be examined using the TR-ARPES. This should have temporal resolution to catch the distribution before electron momentum relaxation smears it, which is $\gtrsim10$ fs. Other phenomena are electron currents and a charge transfer occurring during and after the pulse duration due to the asymmetric electron CB distributions characteristic of the topological resonance.  These can be measured in experiments similar to those of Refs.\ \onlinecite{Higuchi_Hommelhoff_et_al_Nature_2017_Currents_in_Graphene, Hommelhoff_et_al_PhysRevLett.121_2018_Coherent}. 

The topological resonance induces a high CB population and valley polarization. The latter breaks the time reversal symmetry and causes effects similar to those in a very high magnetic field. This will lead, in particular, to anomalous (i.e., without the presence of magnetic field) Hall effect, which can be probed, for instance, by a strong THz radiation similar to Ref.\ \onlinecite{Huber_et_al_s41586-018-0013-6_Nature_2018_Valleytronics} where strong THz pulses were used to control high-harmonic generation in a two-dimensional material.

Finally, a promising perspective application of the topological resonance is an ultimately fast optical memory. The topological resonance allows one to record a reciprocal-space texture during the fundamentally shortest time -- just one optical oscillation, i.e., one or a few fs. This texture can be read out by another single-oscillation pulse during the same ultrashort time. However, the recorded information will live as the CB electron population during a much longer time, likely on the nanosecond scale, which is determined by the CB$\to$VB spontaneous transitions.

\begin{acknowledgments}
Major funding was provided by Grant No. DE-FG02-11ER46789 from the Materials Sciences and Engineering Division of the Office of the Basic Energy Sciences, Office of Science, U.S. Department of Energy. Numerical simulations have been performed using support by
Grant No. DE-FG02-01ER15213 from the Chemical Sciences, Biosciences and Geosciences Division, Office of Basic Energy Sciences, Office of Science, US Department of Energy. The work of V.A. was supported by NSF EFRI NewLAW Grant EFMA-17 41691. Support for S.A.O.M. came from a MURI Grant No. FA9550-15-1-0037 from the US Air Force of Scientific Research.
\end{acknowledgments}
 

%

\end{document}